\documentclass[11pt,preprint]{aastex}

\newcommand{\etal}{{\it et~al.}}
\newcommand{\ltsimeq}{\raisebox{-0.6ex}{$\,\stackrel{\raisebox{-.2ex}{$\textstyle <$}}{\sim}\,$}}

\begin{document}

\title{Main Belt Asteroids with WISE/NEOWISE I:\\Preliminary Albedos and Diameters}

\author{Joseph R. Masiero\altaffilmark{1}, A. K. Mainzer\altaffilmark{1}, T. Grav\altaffilmark{2}, J. M. Bauer\altaffilmark{1,3}, R. M. Cutri\altaffilmark{3}, J. Dailey\altaffilmark{3}, P. R. M. Eisenhardt\altaffilmark{1}, R. S. McMillan\altaffilmark{4}, T. B. Spahr\altaffilmark{5}, M. F. Skrutskie\altaffilmark{6}, D. Tholen\altaffilmark{7}, R. G. Walker\altaffilmark{8}, E. L. Wright\altaffilmark{9}, E. DeBaun\altaffilmark{1,10}, D. Elsbury\altaffilmark{1,11}, T. Gautier IV\altaffilmark{1,12}, S. Gomillion\altaffilmark{1,13}, A. Wilkins\altaffilmark{1,14} }

\altaffiltext{1}{Jet Propulsion Laboratory/California Institute of Technology, 4800 Oak Grove Dr., MS 321-520, Pasadena, CA 91109, USA, {\it Joseph.Masiero@jpl.nasa.gov}}
\altaffiltext{2}{Johns Hopkins University, Baltimore, MD 21218 USA}
\altaffiltext{3}{Infrared Processing and Analysis Center, California Institute of Technology, Pasadena, CA 91125 USA}
\altaffiltext{4}{Lunar and Planetary Laboratory, University of Arizona, 1629 East University Blvd, Kuiper Space Science Bldg. \#92, Tucson, AZ 85721-0092 USA}
\altaffiltext{5}{Minor Planet Center, Harvard-Smithsonian for Astrophysics, 60 Garden Street, Cambridge, MA 02138 USA}
\altaffiltext{6}{Department of Astronomy, P.O. Box 3818, University of Virginia, Charlottesville, VA 22903-0818 USA}
\altaffiltext{7}{Institute for Astronomy, University of Hawaii, Honolulu, HI 96822 USA}
\altaffiltext{8}{Monterey Institute for Research in Astronomy, Monterey, CA USA}
\altaffiltext{9}{UCLA Astronomy, PO Box 91547, Los Angeles, CA 90095-1547 USA}
\altaffiltext{10}{Dartmouth College, Hanover, NH 03755 USA}
\altaffiltext{11}{Notre Dame High School, 13645 Riverside Dr, Sherman Oaks, CA 91423 USA}
\altaffiltext{12}{Flintridge Preparatory School, 4543 Crown Ave, La Canada, CA 91101 USA}
\altaffiltext{13}{Embry-Riddle Aeronautical University, 600 S. Clyde Morris Blvd, Daytona Beach, FL 32114, USA}
\altaffiltext{14}{University of Maryland, College Park, MD 20742, USA}

\begin{abstract}

We present initial results from the Wide-field Infrared Survey
Explorer (WISE), a four-band all-sky thermal infrared survey that
produces data well suited to measuring the physical properties of
asteroids, and the NEOWISE enhancement to the WISE mission allowing
for detailed study of Solar system objects.  Using a NEATM thermal
model fitting routine we compute diameters for over 100,000 Main Belt
asteroids from their IR thermal flux, with errors better than $10\%$.
We then incorporate literature values of visible measurements (in the
form of the H absolute magnitude) to determine albedos.  Using these
data we investigate the albedo and diameter distributions of the Main
Belt.  As observed previously, we find a change in the average albedo
when comparing the inner, middle, and outer portions of the Main Belt.
We also confirm that the albedo distribution of each region is
strongly bimodal.  We observe groupings of objects with similar
albedos in regions of the Main Belt associated with dynamical breakup
families.  Asteroid families typically show a characteristic albedo
for all members, but there are notable exceptions to this.  This paper
is the first look at the Main Belt asteroids in the WISE data, and
only represents the preliminary, observed raw size and albedo
distributions for the populations considered.  These distributions are
subject to survey biases inherent to the NEOWISE dataset and cannot
yet be interpreted as describing the true populations; the debiased
size and albedo distributions will be the subject of the next paper in
this series.

\end{abstract}

\section{Introduction}

Since the discovery of (1) Ceres in 1801 \citep{piazzi} the majority
of observations of asteroids and other minor planets have been
conducted in visible wavelengths.  While visible light can provide
very accurate positions, the interpretation of photometry is rendered
ambiguous by the dependence of the observed flux on both the size and
albedo of the asteroid.  This relationship is described by the
equation:
\[D=\frac{1329}{\sqrt{p_V}}10^{-H/5}\] 
\citep[see][for an overview and references therein for its
  derivation]{harrisAIII}, where $D$ is the diameter in kilometers,
$p_V$ is the visible geometric albedo, and $H$ is the absolute
magnitude which is defined as the apparent magnitude the body would
have $1~$AU from the Sun, $1~$AU from the observer, and at $0^\circ$
phase angle.  Albedos of Solar system objects are observed to vary
from only a few percent up to nearly $100\%$ percent for icy surfaces,
causing diameters inferred from visible data alone to have nearly an
order of magnitude uncertainty.  Multiwavelength visible surveys have
found statistical correlations between the albedos and the visible
colors of asteroids \citep[e.g.][]{ivezic01}, however the accuracy of
the mapping between these two properties for individual objects is
currently being examined \citep{mainzer11tax}.  Using an independent
method of measurement for either the albedo or the diameter allows for
the unique solution of the other, given a visible $H$ value.

Accurate diameters and albedos for a large number of asteroids enable
a number of important areas of research into the history and formation
of the Solar system.  Diameter measurements of asteroids based on
infrared flux allow us to quantify the size-frequency distribution
(SFD) of the bodies in a way independent of assumptions about the
translation from H magnitude to diameter that are typically required.
Both the initial accretion and formation process of asteroids as well
as the subsequent collisional and orbital evolution affect the current
Main Belt SFD, and a well-measured SFD will allow us to put
constraints on these processes.  The albedo of an asteroid, meanwhile,
is a strong function of its composition.  Compositional gradients for
the Main Belt have been shown in the past from both infrared and
spectral surveys \citep[e.g.][]{zellner79,gradie82,tedesco02,bus02},
but only for a limited number of objects.  A large survey conducted in
mid-infrared wavelengths will allow Main Belt albedos and diameters to
be produced with good accuracy; this in turn will allow us to study
the compositional gradient of the Solar System and may ultimately
allow us to set constraints on any major planetary migration that may
have occurred.

Recent work has attempted to understand the SFD of asteroids as it
relates to the impact physics dominating the evolution of the Main
Belt with the aid of numerical simulations.  \citet{obrien03}
analytically calculated the behavior of a population in a steady-state
collisionally dominated regime to determine how the slope of a
population's SFD behaved as a function of the mechanical strength of
the material composing the body.  They find that while a strength-less
regime yields a nominal power-law slope of $a=-3.5$, including the
strength of the body can vary the result over a range of slopes
depending on the specific circumstances.  Expanding on this,
\citet{durda07} simulated a variety of impacts with numerical
hydrocodes to look for the effect of internal structure, impact angle,
impact velocity, and impactor-target size ratio on the resultant SFD
of the shattered products.  These authors have also attempted to match
their numerical results to measured SFDs for asteroid families to
determine the initial size of the parent body, however the data used
have depended on assumptions for the albedos of the asteroids.  For
the larger Main Belt population, \citet{bottke05a} have modeled the
evolution of various initial size distributions to determine what the
current MBA SFD can indicate about the SFD present at the formation of
the asteroids.  Their results show a clear peak in formation SFD, with
few objects smaller than $D\sim100~$km forming directly from the
protosolar nebula, however these results are also based on diameters
determined from assumed albedos.  Similarly, \citet{bottke05b} have
investigated the sources and sinks of excited MBAs and the effect
orbital evolution has the SFD of the Main Belt.

Studies of asteroid albedo distributions have in the past been limited
by small data sets.  Early measurements were made using thermal
infrared detectors on ground-based telescopes to determine radiometric
diameters and thus albedos for tens of objects
\citep[e.g.][]{morrison74}.  The {\it InfraRed Astronomical Satellite}
(IRAS) revolutionized studies of physical properties of the Main
Belt by observing over $2000$ asteroids in the thermal infrared,
determining albedos and diameters for these bodies in a uniform way
\citep{tedesco02,tedescoPDS}.  To date, IRAS represents the
largest and most complete survey of asteroid albedos in the
literature.

If the sizes and albedos of the members of an asteroid family are
known, it is possible to use their orbital evolution to study their
age.  The Yarkovsky effect \citep[see][for a review of the
  subject]{bottke06} occurs when incident optical light is absorbed
and re-emitted as thermal infrared photons in a different direction,
usually due to the rotation of the body.  This difference in momentum
causes changes in the orbit of the body over long time intervals.
\citet{nesvorny04} use this effect to refine predictions for the age
of the Karin family based on backwards integration of orbital
parameters.  As Karin is one of the youngest families
\citep{nesvorny02} this effect becomes increasingly important for
accurate age-dating of families the older they are.  The accuracy of
this technique, however, depends on knowing sizes and albedos of the
objects to a high degree of certainty.

The distribution of albedos in the Main Belt as a whole and broken
down by region can provide us a window into the changing chemical and
mineralogical processes active in different regions of the early Solar
system.  Although the current understanding of the history of the Main
Belt has the asteroids forming in or near their current locations, new
theories are being proposed that the Main Belt may in fact be the
result of a mixing of two distinct populations from different regions
of the Solar system.  Migrations of the giant planets may have both
cleared many of the objects that initially formed in the Main Belt
region and repopulated this area with objects from beyond the ``snow
line'' \citep{morbi10,walsh11}.  We then might expect the Main Belt to
be composed of two overlapping populations, one having formed in a
volatile-poor region of the protosolar disk and one forming in a
volatile rich area, though the latter population would lose any
surface volatiles over the age of the Solar system.  Under this
scenario albedo may be able to trace dynamical evolution as well as
chemical processing.

There are very few ways to measure the albedo of an asteroid directly.
Observations taken {\it in situ} by spacecraft can be used to measure
both the absolute albedo and its variation across a body's surface
\citep[e.g.][]{howett10}, as well as its diameter, though only a
handful of objects have been visited by spacecraft.  Imaging
polarimetry can also be used to determine the integrated surface
albedo for a number of objects \citep{cellino99}, but appropriately
calibrated instrumentation is uncommon.  Similarly, direct
measurements of asteroid diameters can come from a variety of
techniques.  Resolved imaging of an asteroid provides the simplest
method of size determination, however the use of e.g. {\it HST} or
Keck adaptive optics (AO) only allows the largest few asteroids to be
resolved \citep[e.g.][]{schmidt09,li10}.  Asteroid occultations of
background stars also provide a robust method of diameter measurement
\citep{shevchenko06}, however the logistical constraints of
occultation events make obtaining a large sample difficult, and shape
measurements are only instantaneous projections.  Radar measurements
can provide precise distances, rotation rates, projected profiles, and
with sufficient data 3D shape models \citep{ostroAIII}, but returned
fluxes fall off quickly with distance which limits the number of
objects observable with this technique.

Indirect measurements of asteroid diameters and albedos can provide a
wealth of information both for individual objects and the population
as a whole.  These techniques typically can observe a large number of
objects in a relatively short period of time, provide uniform data for
an entire population, and have understood biases that allow for
determination of the true, underlying distributions.  Careful
calibration of indirect measurements is required, but once established
these techniques can provide highly accurate measurements of asteroid
physical properties for a large number of bodies.

For objects that have known orbits, measurement of the infrared flux
emitted from the surface can be used to constrain the diameter of the
body (see \S\ref{sec.models} for a discussion of the method used
here).  Infrared imaging can be accomplished rapidly when integrated
in an all-sky survey.  Thermal infrared measurements of a large sample
of asteroids represent the best way to determine robust diameters
rapidly for many thousands of objects.  In this paper, we present
preliminary results from the {\it Wide-field Infrared Survey Explorer}
(WISE) space telescope, the next-generation all-sky infrared survey,
focusing here on the cryogenic observations of Main Belt asteroids.

\section{Observations}
\subsection{WISE and NEOWISE}
\label{sec.wise}
Launched on 2009 December 14, WISE is a thermal infrared space
telescope that performed an all-sky survey from 2010 January 14 until
it exhausted the telescope-cooling hydrogen ice on 2010 August 5.  The
survey continued during the warmup of this secondary tank with limited
sensitivity in the longer wavelengths until the primary coolant,
responsible for maintaining detector temperature, exhausted 2010
October 1.  WISE subsequently entered a Post-Cryogenic Mission to
complete the survey of the largest MBAs, continue discovering new
NEOs, and complete a second-pass survey of the inertial sky in the two
shortest wavelengths.  During its cryogenic mission, WISE imaged the
sky in four infrared wavelengths simultaneously using dichroic
beamsplitters to produce co-boresight images with band centers at
$3.4~\mu$m, $4.6~\mu$m, $12~\mu$m and $22~\mu$m (W1, W2, W3, and W4,
respectively).  The latter two bands are particularly important for
Solar system studies as the dominant amount of the flux received from
asteroids is from thermal emission peaking at these wavelengths.
First-pass calibration of the WISE data was tuned to the fully
cryogenic mission and while the final calibration currently being
undertaken will finalize measurements obtained during each warm-up
stage, we restrict our current analysis to objects observed during
this fully cryogenic stage.  Pre-launch descriptions of WISE were
given by \citet{mainzer06} and \citet{liu08}, while post-launch
overviews, including initial calibrations and color corrections, are
presented by \citet{wright10} and \citet{mainzer11cal}.

The WISE survey follows a continuous scan along lines of ecliptic
longitude at a solar elongation of $\sim90^\circ$ as the spacecraft
orbits above the terminator of the Earth.  An oscillating scan mirror
compensates for this motion, providing stable images with effective
integration times of $8.8~$seconds.  The positions of all known minor
planets are propagated to the time of each observation and checked
against all transient sources in the field of view, recording the
appropriate calibrated magnitudes if observed.  In this way, thermal
measurements of each minor planet cataloged by the Minor Planet Center
(MPC)\footnote{see the MPCORB.DAT file available here: {\it
    http://www.minorplanetcenter.net/iau/MPCORB.html}} can be found.
In addition to previously known objects, the WISE processing pipeline
includes the capability to discover new objects via the NEOWISE
enhancement \citep{mainzer11}.  While comprehensive followup of new
potential near-Earth objects (NEOs) has been one of the priorities of
the WISE team \citep{mainzer11neo}, the large number of Main Belt
discoveries ($>34,000$) has prohibited sufficient immediate
ground-based followup.  Large-scale asteroid surveys such as
Spacewatch and the Catalina Sky Survey have already provided
serendipitous followup of many of the new WISE MBAs and will continue
to do so, while future surveys are expected to recover the remainder.
As described in \citet{wright10} and \citet{mainzer11}, the WISE
cadence resulted in an average of 10-12 observations of each minor
planet detected over $\sim36~$hours.  For a subset of the Main Belt,
predominantly in the outer regions, observations were obtained at two
or more epochs depending on the relative motion of the object and
Earth around the Sun.

\subsection{Data and calibration}
\begin{figure}[ht]
\begin{center}
\includegraphics[scale=0.4]{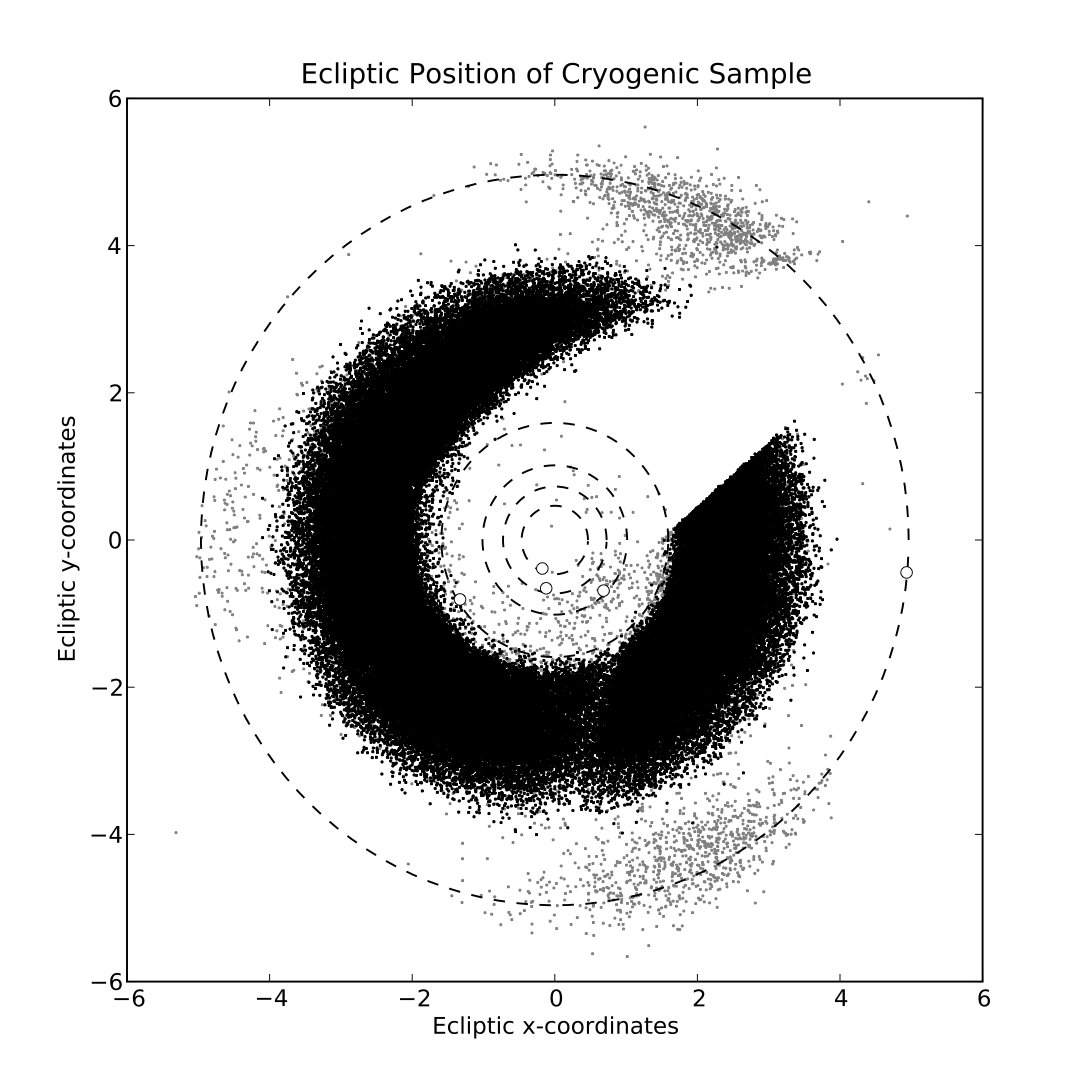} \protect\caption{Top-down view
  of the inner Solar system showing the location of all objects
  observed during the fully-cryogenic mission.  Positions were
  propagated to 2010 August 5, the date of the exhaustion of coolant
  from the secondary tank.  Black points indicate MBAs while grey
  points are all other Solar system objects.  Axes' units are AU.}
\label{fig.pacman}
\end{center}
\end{figure}

In this paper, we consider only those Main Belt asteroids detected by
NEOWISE/WMOPS during the cryogenic portion of the mission, shown in a
top-down view of the inner Solar system in Fig~\ref{fig.pacman}.  This
consists of $129,750$ unique objects.  We obtained our data used for
fitting in a method identical to the one described in
\citet{mainzer11cal} and \citet{mainzer11neo}, though tuned for MBAs.
Specifically, we queried the Minor Planet Center observation
file\footnote{{\it http://www.minorplanetcenter.net/iau/ECS/MPCAT-OBS/MPCAT-OBS.html}}
for all observations submitted from WISE (observatory code C51).  We
then used the resultant RA-Dec-time values as input for a query of the
WISE individual exposure archive, the ``Level 1b'' data, through the
Gator tool provided by the InfraRed Science Archive
(IRSA\footnote{{\it
    http://irsa.ipac.caltech.edu/applications/Gator/}}).  To ensure
that only the observations of the moving objects were returned, we
restricted our search radius to 0.3 arcsecs from the position and 2
seconds from the time obtained for the detection from the MPC.
Additionally, we set a constraint of JD$<2455413.5$ to ensure that only
data during the fully cryogenic mission was used for this initial
survey.  This method of data acquisition has the benefit of ensuring
all observations used have been vetted both internally by the WISE
data pipeline and again by the MPC.

All data were processed using the first pass version of the WISE
pipeline, which computed dark current/sky offset levels, flagged
instrumental artifacts such as latent images and diffraction spikes
and performed linearity compensation.  Only observations with an
artifact flag cc\_flags$=0$ or $p$ in a band were accepted: a value of
$0$ indicates no evidence of artifact was found by the pipeline, while
$p$ indicates the possibility of contamination by a latent image.  As
discussed in \citet{mainzer11cal} we find that the pipeline was overly
conservative in artifact flagging and cc\_flags$=p$ values have similar
fluxes to cc\_flags$=0$ detections while increasing the number of
usable observations by $\sim20\%$.  The ph\_qual flag was required to
have a value of A, B or C to again ensure only valid detections were
used.  Non-linearity and saturation are a particular concern for the
brightest MBAs, especially in bands W3 and W4.  The WISE data
reduction pipeline applies a non-linearity and saturation correction
for all observations brighter than the threshold of $W1=7.8$~mag,
$W2=6.5$~mag, $W3=3.6$~mag, $W4=-0.6$~mag.  Objects with magnitudes
brighter than $W3=4$ and $W4=3$ were assigned errors of $0.2~$mag to
account for the change in the point-spread function for very bright
objects, and a linear correction to the magnitudes of sources with
$-2<W3<4$ was applied \citep{mainzer11cal,cutri11}.  Following those
authors, we did not use objects brighter than $W3=-2$ and $W4=-6$ for
thermal modeling.

Each object was required to have been observed at least 3 times in one
WISE band with magnitude error $\sigma_{mag}\le0.25$ to undergo
thermal modeling, as a precaution against contamination by spurious
sources (e.g. background noise, cosmic rays, stars, etc.).  For
multiple-band thermal models we required other bands to have at least
$40\%$ of the detection rate of the band with the largest number of
detections, usually W3 for MBAs.  To reduce the possibility of
confusion with inertially fixed sources such as stars and galaxies, we
searched each position retrieved from the Level 1b catalog in the
Daily and Atlas Coadded Catalogs (also served by IRSA) within 6.5
arcsecs, equivalent to the W1, W2 and W3 beam sizes.  These searches
looked for sources that appeared at least twice and in at least $30\%$
of the images covering that location.  Any sources returned from these
searches were considered to be inertial, which could contaminate the
observation of the asteroid at that position.  Thus that asteroid
detection was discarded from the thermal modeling routine.

\begin{figure}[ht]
\begin{center}
\includegraphics[scale=0.8]{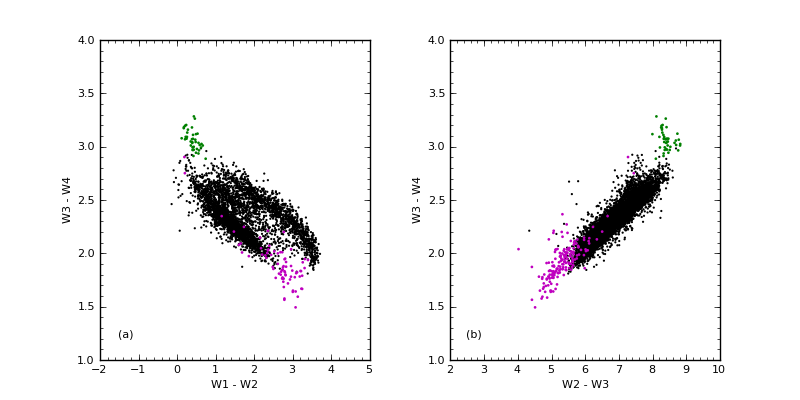}
\protect\caption{(a)W1-W2 vs. W3-W4 colors for all MBAs (black),
  Trojans (green), and NEOs (magenta); (b) W2-W3 vs. W3-W4 colors for
  these same populations.}
\label{fig.colors}
\end{center}
\end{figure}

\begin{figure}[ht]
\begin{center}
\includegraphics[scale=0.7]{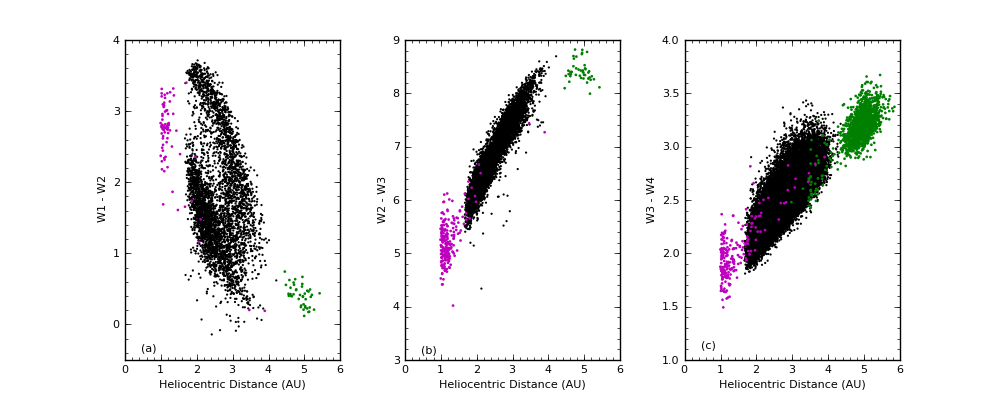}
\protect\caption{(a)W1-W2 vs. heliocentric distance for all MBAs
  (black), Trojans (green), and NEOs (magenta); (b) as in (a) but for
  W2-W3 color; (c) as in (a) but for W3-W4 color.}
\label{fig.colordist}
\end{center}
\end{figure}

\begin{figure}[ht]
\begin{center}
\includegraphics[scale=0.8]{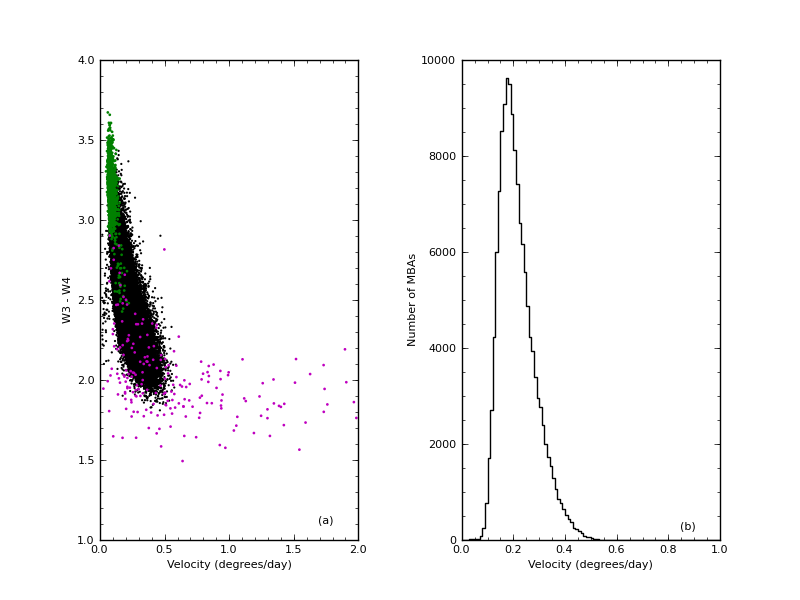}
\protect\caption{(a)W3-W4 color vs. sky-plane velocity for all MBAs
  (black), Trojans (green), and NEOs (magenta); (b) differential
  distribution of sky-plane velocities for all MBAs.}
\label{fig.velocity}
\end{center}
\end{figure}

In Fig~\ref{fig.colors} we show the mean colors of inner Solar system
objects as observed during the cryogenic phase of the WISE mission
with the MBAs highlighted in black \citep[see][for discussions on the
  other populations shown in this figure]{mainzer11neo,grav11}.  The
bifurcation in the W1-W2 color observed for the MBAs traces the two
dominant albedo groupings in the Main Belt (see below for further
discussion).  The MBAs span a wide range of colors and are bounded by
the NEO and Jupiter Trojan populations.  Note that MBAs occupying
color-space typically associated with one of the other populations are
candidates for objects that may have been misidentified as MBAs during
preliminary orbit fitting.  We show in Fig~\ref{fig.colordist} the
color for each object as a function of heliocentric distance at the
time of observation.  Color, especially in bands dominated by thermal
emission, is a strong function of temperature of the body and thus
distance from the Sun.  Figure~\ref{fig.velocity} shows the sky-plane
velocity for each object compared with its W3-W4 color.  The mean
sky-plane velocity for MBAs is $0.2~$deg/day; objects with velocities
significantly larger than this are candidates for NEOs misclassified
as MBAs, requiring further followup.  Future work will combine color
and sky-plane velocity cuts to distinguish MBAs with poorly known
orbits from NEOs and Trojans.

\clearpage

\section{Diameter and albedo determination through thermal modeling}
\label{sec.models}
In contrast to the visible flux received from an asteroid, which is
reflected sunlight, the mid-IR flux beyond $\sim6~\mu$m from an object
in the Main Belt is almost completely thermal emission from that body.
For an object with an established orbit, the phase angle and distances
to the asteroid from the Earth and the Sun are well known, thus the
observed thermal flux can be converted into a total emitted flux at
the surface.  By making assumptions for some physical surface
properties, a diameter can be computed from a single band detection.
When simultaneous thermal measurements at multiple wavelengths are
available the beaming parameter of the surface material may also be
fit to the data.  The beaming parameter ($\eta$) represents the
deviation of the thermal emission from that of a smooth non-rotating
sphere due to rotation and surface roughness, and is used to
consolidate the uncertainty in the values of the surface thermal
properties, including emissivity.  When visible-light data are
additionally available, the visible albedo ($p_V$) becomes a free
parameter that can now also be fit using the two data sets in
conjunction.  Further data such as W1 or W2 measurements that are
dominated by reflected sunlight allow us to derive an independent
measure of the ratio of visible to NIR reflectance as well.

In order to analyze the thermal infrared asteroid measurements from
the IRAS satellite, \citet{STM} developed the ``Standard Thermal
Model'' (STM) for asteroids, calibrated against measurements from (1)
Ceres and (2) Pallas.  In this model, the beaming parameter was held
constant to $\eta=0.756$ based on the ground-truth occultation
observations of the calibrator asteroids.  Subsequent work indicated
that this model might not be appropriate for smaller asteroids, thus
\citet{NEATM} modified STM to a form appropriate for a ``Near-Earth
Asteroid Thermal Model'' (NEATM), where $\eta$ is allowed to vary.
While designed specifically to account for the breakdown of STM when
considering NEOs, NEATM can be applied quite readily to a wide range
of bodies in the Solar system \citep[e.g.][]{mueller10,ryan10}.
\citet{wright07} compared NEATM with a full thermophysical model of a
cratered surface and found that for low phases both models produce
consistent results.  \citet{FRM} present an investigation of objects
with a beaming parameter at the theoretical maximum of $\eta=\pi$
which occurs for a body rotating so quickly it is latitudinally
isothermal.

We have performed preliminary thermal modeling of MBAs based on the
WISE First-Pass Data Processing Pipeline described above, covering
observations taken during the cryogenic phase of the mission.  We
modeled each object as a non-rotating sphere with triangular facets
and variable diameter, beaming parameter, visible albedo, and NIR
reflectance ratio as appropriate for the data.  Relative distances and
phase angles were computed for each measurement to ensure changing
distances do not bias the resultant fits.  Although we do not expect
all, or even most, asteroids to have a spherical shape, our
observations covering $\sim36~$hours smooth out rotation effects and
allow us to determine the effective diameter of a spherical body with
the same physical properties.  Long period rotators ($P\sim$days) with
large amplitudes, e.g. binary asteroids with mass fraction $\sim1$,
will have poor fits resulting in a moderate mis-measurement of albedo
and diameter.  Future work will address the light curve component of
our data set to determine the minimum and maximum sizes of our targets
in order to estimate first-order shape models as well as estimates of
the fraction of binaries in the Main Belt.

At each instance, the temperature on every facet was computed and
color corrected based on the values in \citet{wright10}.  The emitted
thermal flux for each facet was computed with the NEATM model and
night-side facets were assumed to contribute zero flux.  A reflected
light model was used to determine the reflected component in each band
for all illuminated facets visible to WISE at the time of observation;
this was most important for W1 and W2.  The model reflected and
emitted light was summed for all facets and converted to a model
magnitude using the Jansky flux of a zero magnitude source provided in
\citet{wright10} and modified according to the text for red sources:
$306.681~$Jy for W1, $170.663~$Jy for W2, $31.3684~$Jy for W3 and
$7.9525~$Jy for W4.  Note that the modifications in the W3 and W4
zeropoints are the result of adjusting the central wavelengths of
these bands to $\lambda_{0,W3}=11.0984~\mu$m and
$\lambda_{0,W4}=22.6405~\mu$m to correct the discrepancy observed in
the calibration tests between blue-spectrum and red-spectrum objects.
These model magnitudes were then compared with the measured
magnitudes, and the model was iterated through a least-squares fitting
routine until the best fit was found.

For objects with two thermally dominated bands, the beaming parameter
was allowed to vary, while for those with only one thermal band we
used a fixed value of $\eta=1.0$, based on the peak of the $\eta$
distribution of MBAs that were fit with a variable beaming parameter
(see below).  For objects where no detected band was dominated
($>75\%$) by reflected light we assumed a NIR reflectance ratio of
$1.5$ as was found for MBAs with fitted NIR reflectances (see below).
These objects typically were not detected in W1 and either were not
detected or had both thermal emission and reflected light in W2.

Monte Carlo simulations were performed in each case to determine the
errors on all variable parameters.  For high S/N cases where the
quoted error on the measured magnitude only represented the
statistical error, we set a floor of $\sigma_{mag}=0.03~$,
representing the absolute error on the photometry \citep{wright10}.
For all objects we assumed that the emissivity $\epsilon=0.9$ and we
assumed that the magnitude-phase slope parameter
\citep[c.f.][]{bowell89} was $G=0.15$ unless otherwise given by the
MPC or in the Lightcurve Database \citep[LCDB\footnote{\it
    http://www.minorplanet.info/lightcurvedatabase.html},][]{warner09a}.
The quoted errors on the modeled parameters are equal to the weighted
standard deviation of all Monte Carlo trial values.  For objects with
fixed beaming parameter, an error of $\sigma_\eta=0.2$ was assumed to
allow for proper error determination of derived parameters based on
the mean and standard deviation of all best-fitting beaming parameters
for objects with fitted values (see \S\ref{sec.etavar}).  Similarly, for
objects with fixed IR reflectance ratios, we assume an error bar of
$\sigma_{ratio}=0.5$ based on the mean and standard deviation of
objects with fitted IR reflectance ratios (see \S\ref{sec.ratio}).  We
note that as the flux calibrations presented by \citet{mainzer11cal}
set a limit on the computed diameter accuracy for sources in the WISE
data of $\sigma_D=10\%$.  This error implies a minimum fractional
error for albedo of $\sigma_{p_V} = 20\% \times p_V$ assuming a
perfect H magnitude.  These values are in addition to any Poissonian
error inherent to the observations, though for most of the objects
presented here the calibration errors dominate our solutions.

In total $129,750$ Main Belt asteroids, selected from the cryogenic
phase of the survey, had sufficient number and quality of detections
to allow us to perform thermal modeling and determine their effective
diameter.  Of these, $17,482$ objects had orbital arcs shorter than
$30$ days; as such their orbits have a larger uncertainty than the
rest of the population, which corresponds to uncertainty in their
geocentric and heliocentric distances, which will naturally increase
the error on their calculated diameters.  Additionally, $112,265$
objects also had available optical data allowing us to calculate
albedo as well as diameter.  Both of these latter two populations are
changing continuously, as ground-based surveys submit serendipidous
visible observations of NEOWISE-discovered asteroids (allowing us to
fit albedos and allowing the MPC to fit better orbits), and as the MPC
links WISE observations with previous one-night stands and lost
asteroids.

We provide a table of our best fits for Main Belt asteroids from the
Pass 1 processed cryogenic survey data online at: {\it
  http://wise2.ipac.caltech.edu/staff/bauer/NEOWISE\_pass1/}.  This
table contains: the MPC-packed format name of the object; the $H$ and
$G$ values used; the diameter, albedo, beaming parameter, and infrared
albedo as well as associated error bars; the number of observations in
each WISE band that were used for fitting; and the mean modified
Julian date of the observations.  For objects observed at multiple
epochs with fits consistent across all observations, each epoch is
presented as a separate row in the table.  As discussed in
\S\ref{sec.unusual} objects where multiple epochs were forced to fit
to a single model because the separate fits were divergent are
presented as a single row.  Objects without optical data at the time
of publication have ``nan'' (``not a number'') values for absolute
magnitude and albedo.  Similarly, objects for which an infrared albedo
could not be fit nor had literature optical data that could be used
with an assumed reflectance ratio to estimate an infrared albedo (see
\S\ref{sec.ratio}) have ``nan'' values in this entry.

\section{Preliminary raw size-frequency distribution of MBAs}
\label{sec.sfd}

\begin{figure}[ht]
\begin{center}
\includegraphics[scale=0.75]{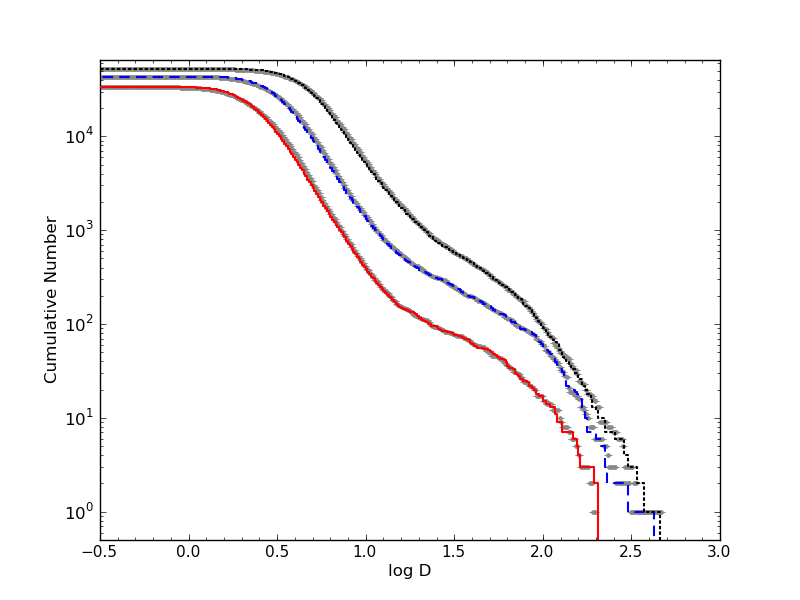}
\protect\caption{Cumulative raw size-frequency distribution of MBAs in
  the inner- (red), middle- (blue), and outer-Main Belt (black).
  Plotted under the distributions are grey points showing the
  Monte Carlo simulation for each data set; the error bars are the
  size of the points.}
\label{fig.dCum}
\end{center}
\end{figure}

Using fluxes from the WISE data, and a faceted NEATM model, we are
able to determine diameters for our observed objects.  In
Fig~\ref{fig.dCum} we show the cumulative preliminary raw
size-frequency distribution (PRSFD) for the three major regions of the
Main Belt: the inner-Main Belt (IMB, those objects with $1.8~$AU
$<a<2.5~$AU), the middle-Main Belt (MMB, objects with $2.5~$AU
$<a<2.82~$AU), and the outer-Main Belt (OMB, objects with $2.82~$AU
$<a<3.6~$AU).  In all cases, perihelion distance was required to be
beyond the orbit of Mars, $q>1.666~$AU.  Also plotted on the
distribution histograms are 100-trial Monte Carlo (MC) simulations of
the diameter distribution including the appropriate measured errors.
This mean MC distribution and associated error are shown as points;
the error bar sizes are smaller than the point size.  We see no
significiant change between the MC distribution from the distribution
of best-fit diameters.  We find the slope of the PRSFD for smaller
objects in all three subpopulations to be consistent with the $a=-2.5$
value found by \citet{gladman09}, however debiasing will be critical
to determining the true value of this slope.  We observe a significant
change in slope for the PRSFD between $15-25~$km, consistent with the
location of the ``kink'' seen in the debiased $H$ distribution by
\citet{jedicke98}.  A debiasing campaign of the observed population
currently underway will allow us to explore the true SFDs of the
populations and will be discussed in future work.

\section{Variable beaming parameters}
\label{sec.etavar}

\begin{figure}[ht]
\begin{center}
\includegraphics[scale=0.75]{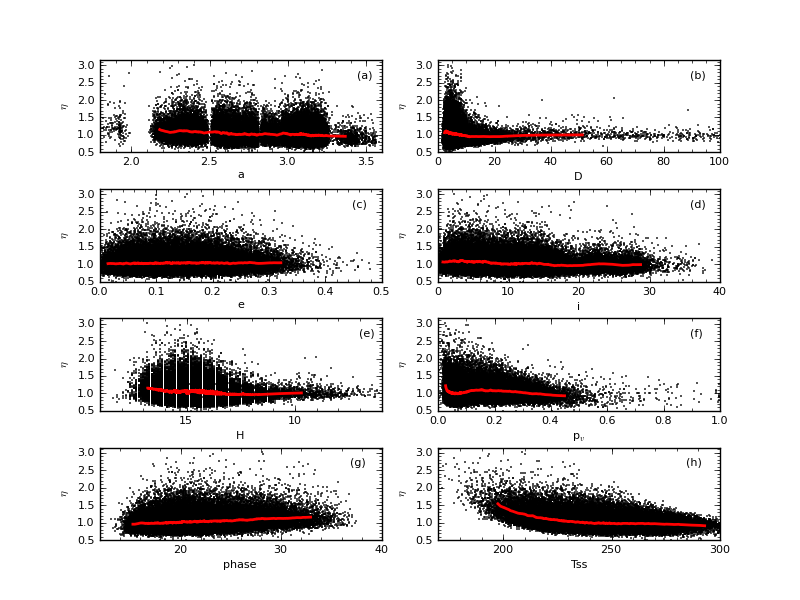} \protect\caption{Beaming
  parameter for all objects with fitted values, compared to (a)
  semi-major axis, (b) diameter, (c) eccentricity, (d) inclination,
  (e) absolute magnitude, (f) albedo, (g) phase and (h) subsolar
  temperature.  The thick red line shows the running average for
  $1000$ object-wide bins stepped by $100$ objects.  The picket-fence
  effect in the absolute magnitude is an artifact of the $0.1~$mag
  reported precision of $H$ for most objects.}
\label{fig.etascat}
\end{center}
\end{figure}

For objects with detections in at least two thermal bands, we allow
the beaming parameter to vary during the thermal model fitting.  We
are able to fit beaming parameters for $66,406$ MBAs.  We find a wide
range of best-fit beaming parameters between the theoretical limits of
$0.3$ and $\pi$, with a peak value of $\eta=1.0$ and standard
deviation of $\sigma_\eta=0.2$.  In Fig~\ref{fig.etascat} we show the
beaming parameter found for all objects with fitted values as a
function of a variety of orbital and physical parameters.  Beaming
parameter has a weak dependence on semi-major axis
(Fig~\ref{fig.etascat}a), an effect that is more pronounced for the
relationship with phase angle (Fig~\ref{fig.etascat}g).  Note that due
to the constraints of the pointing of WISE over the course of the
survey to solar elongations $\approx90^\circ$, phase angle,
heliocentric distance and geocentric distance are strongly correlated,
though ecliptic latitude of the observations as well small changes in
the exact pointing over the survey weaken this relation.
\citet{wright07} has shown that for the phase angles we typically
observe MBAs at ($14^\circ\ltsimeq\alpha\ltsimeq27^\circ$) the differences
in calculated diameter and beaming parameter between NEATM and more
realistic thermophysical models is minimal over a large range of
observing geometries.  From the raw distribution, the beaming
parameter shows no dependence on size (Fig~\ref{fig.etascat}b),
eccentricity (Fig~\ref{fig.etascat}c), inclination
(Fig~\ref{fig.etascat}d) or absolute magnitude
(Fig~\ref{fig.etascat}e).

We find a best-fit linear relation to the running average of beaming
parameter as a function of phase of:
\[\eta = (0.79\pm0.01) + \alpha (0.011+0.001)\]
where $\eta$ is the beaming parameter and $\alpha$ is the phase angle
in degrees.  This is consistent with the results found in
\citet{mainzer11neo} for the NEOs from WISE, but differs significantly
from the results of \citet{wolters09} who found a best-fitting line of
$\eta=1.08 + 0.007\alpha$.  For phase angles within the Main Belt
($14^\circ - 32^\circ$) the average beaming ranges from
$0.94<\eta<1.14$ although the spread around this value is large.  Thus
$\eta_{assumed}=1.0$ for objects with only a single thermal band is a
reasonable assumption for objects in the Main Belt.  Debiasing will
allow us to account for any detection-limit effects that may bias the
selection of objects that have sufficient data for fitting of the
beaming parameter.

Note that while the running average over the beaming parameter shows a
dependence on albedo (Fig~\ref{fig.etascat}f) and subsolar temperature
(Fig~\ref{fig.etascat}h), albedo determinations are very sensitive to
survey biases, both from WISE and optical followup, while subsolar
temperature is a function of beaming parameter as per the equation
from the definition of NEATM in \citet{NEATM}:
\begin{equation}
T_{SS} = [(1-A)S/(\eta\epsilon\sigma)]^{0.25}
\label{eq.Tss}
\end{equation}
(where $T_{SS}$ is the subsolar temperature, $A$ is the Bond albedo,
$S$ the incident Solar flux, $\eta$ the beaming parameter, $\epsilon$
the emissivity, and $\sigma$ the Stefan-Boltzmann constant), and so
this cannot be used as an independent constraint.

\clearpage

\begin{figure}[ht]
\begin{center}
\includegraphics[scale=0.75]{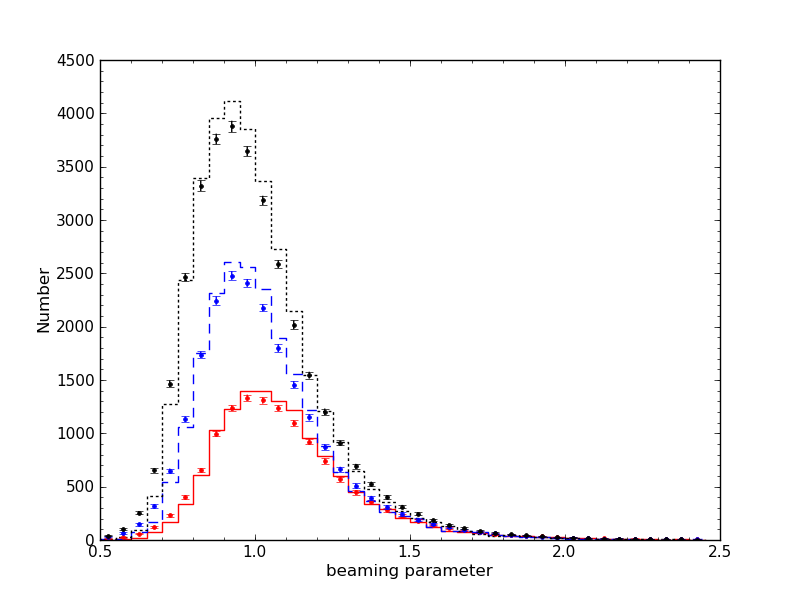}
\protect\caption{Histograms show the beaming parameter distribution
  for the IMB (red solid), MMB (blue dashed) and OMB (black dotted)
  populations.  The points with error bars show the mean Monte Carlo
  distribution and associated error.}
\label{fig.etahist}
\end{center}
\end{figure}

We show in Fig~\ref{fig.etahist} histograms of the beaming parameter
distribution for the inner-, middle-, and outer-Main Belt populations.
Also shown as points are the mean distribution and error from a
100-trial Monte Carlo simulation of the distribution using the error
bar on each fitted beaming parameter.  All populations show longer
tails toward higher values of beaming parameter than toward lower, and
have consistent shapes.  The shift in peak beaming parameter with phase
can be seen as the change in distribution between populations with
different average phase angles.

\section{Preliminary raw albedo distribution of MBAs}

With the inclusion of visible data in our modeling, we determine
albedos as well as diameters for the asteroids discussed here.  We use
the published $H$ and $G$ values for all asteroids, available from the
MPC.  During the confirmation of the calibration of WISE for
asteroids, \citet{mainzer11cal} investigated the need for an offset in
H to account for systematic errors in H values, but found that no
offset was required \citep[c.f.][who found a $0.2~$mag
  shift]{juric02}.  The $H$ magnitudes were assigned a random error of
$\sim0.2~$mag.  We perform Monte Carlo simulations of our visible
light measurements as well as of the thermal measurements to quantify
the error on albedo, however in all cases the minimum error on albedo
will be $20\%$ \citep{mainzer11cal} for objects with optical data and
one good thermal band.  We have sufficient optical data to determine
albedos for $112,265$ MBAs.  Though our individual albedos have large
error bars, the population statistics can still provide us with a
window into the state of the surface composition of the Main Belt.

\begin{figure}[ht]
\begin{center}
\includegraphics[scale=0.75]{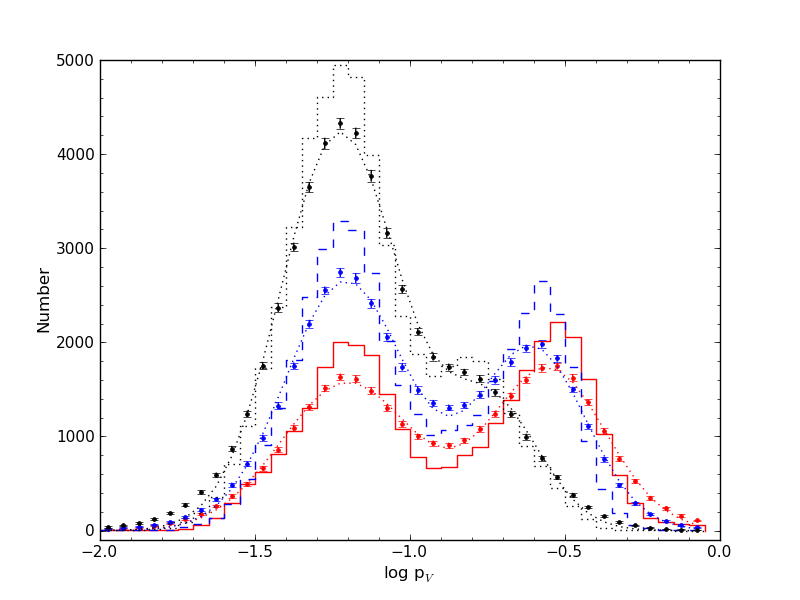}
\protect\caption{Preliminary raw differential albedo
  distributions for all inner-, middle-, and outer-Main Belt
  asteroids, shown as red solid, blue dashed, and black dotted
  histograms respectively.  The points show Monte Carlo simulations of
  the albedos and their error bars, and the smooth curves the best
  fitting double-Gaussian distributions.}
\label{fig.albAll}
\end{center}
\end{figure}

In Fig~\ref{fig.albAll} we show the differential preliminary
raw albedo distribution (PRAD) of all the inner-, middle-, and
outer-Main Belt asteroids in our survey.  We then take the fitted
albedos and their respective error bars and perform a 100-trial MC
simulation of these values to find a mean distribution with errors,
shown as points.  In all cases, the peaks of the distributions broaden
slightly in the MC simulation, which is expected.  In log-albedo-space
the differential distribution is well described by a bimodal Gaussian
distribution.  We show our best-fitting double-Gaussian (fitted to the
mean distribution found through the MC simulations) as the smooth
dotted curve under each set of points.

The bimodality in albedo likely traces the difference between the two major
branches of asteroid composition: the S-type asteroids with high
albedos and the C-type asteroids with low albedos
\citep{chapman75,tedesco89}.  \citet{mainzer11tax} investigate the
specific link between albedo and a variety of taxonomic
classification systems.  As discussed above, however, the PRAD will
naturally include the observational biases of the ground-based
telescopes used to determine the optical magnitudes needed to find the
visible albedo, favoring higher albedo asteroids and over-representing
their contribution to the total population, particularly in the Inner
Main Belt.  Debiasing, currently being undertaken, will allow us to
quantify and remove this effect.

The mean value and width of the Gaussian that best describes the dark
peak of the PRAD for each population is consistent across populations,
with mean albedo $\mu=0.06$ and a dispersion of $\sigma =
^{+.03}_{-.02}$.  Note that the Gaussian error bars on the $\mu$ value
are in log space, and thus asymmetric in native units.  Unlike the
dark asteroids, the bright complex shows a distinct change in the mean
value in the PRAD as one moves out in the Main Belt.  The mean albedo
of the bright peak for the Gaussian describing each population is:
$\mu_{IMB} = 0.28$, $\mu_{MMB} = 0.25$, and $\mu_{OMB} = 0.17$, with
widths of $\sigma_{IMB} = ^{+.13}_{-.09}$, $\sigma_{MMB} =
^{+.11}_{-.08}$, and $\sigma_{OMB} = ^{+.08}_{-.05}$.

The objects discovered by NEOWISE that have not had optical followup
will add a significant bias to the PRAD as these objects are most
likely to be ones missed by optical surveys, i.e. lower albedo
asteroids.  During the cryogenic portion of the survey, NEOWISE
observed $23,616$ previously unknown Main Belt asteroids with data of
sufficient quality for thermal modeling and with orbital arcs longer
than $1$ day (and thus not considered ``one night stand'' observations
by the MPC).  Some $\sim10,000$ additional asteroids were given
temporary designations by the MPC but did not have sufficient arc
length to calculate an orbit.  While they do not have known orbits and
thus can't be classified as members of the Main Belt or not, we can
use this to set an upper limit on the number of lost MBAs.  As these
objects are linked to older precovery data or are serendipitously
followed up our count of discovered objects with computed orbits will
increase.  Out of these discovered objects $19,178$ have optical
photometric data as well as thermal infrared, allowing for albedo
determination.  

It should be noted that while many of the discovered objects have
optical observations, there is a strong bias in favor of recovery of
the highest albedo discoveries by subsequent ground based
observations.  The visible light received from an object is directly
proportional to the albedo of that object, while the thermal infrared
flux is driven by the temperature of the surface, which is only weakly
dependent on albedo as shown in Eq~\ref{eq.Tss}.  As such, optical
surveys show a significant albedo bias towards brighter objects for
both discovery and recovery observations, while the NEOWISE infrared
survey is more sensitive to detection of low albedo objects.  Any
survey will have inherent biases in the data set and it is necessary
to account for them before the true albedo distribution can be
determined for a population.

\begin{figure}[ht]
\begin{center}
\includegraphics[scale=0.75]{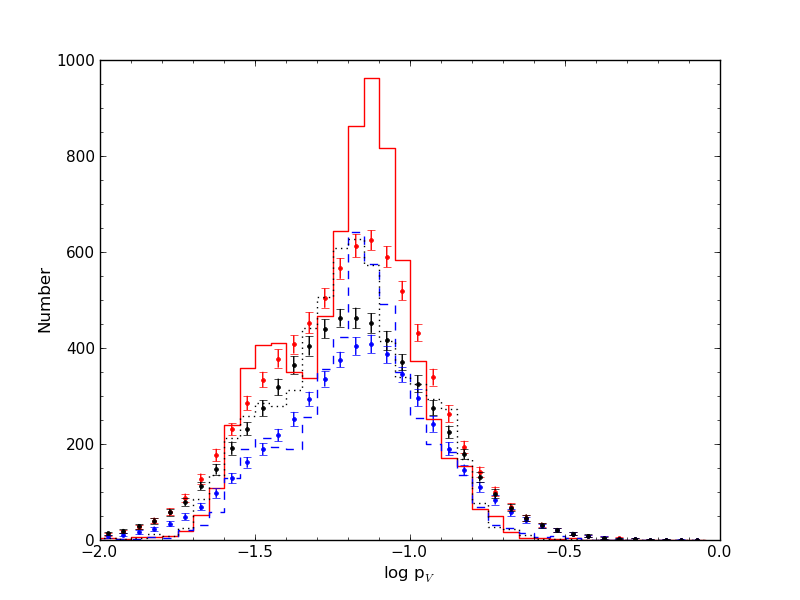}
\protect\caption{The same as Fig~\ref{fig.albAll}, but for only the
  NEOWISE-discovered MBAs that also received optical photometric
  followup.}
\label{fig.discalb}
\end{center}
\end{figure}

\begin{figure}[ht]
\begin{center}
\includegraphics[scale=0.75]{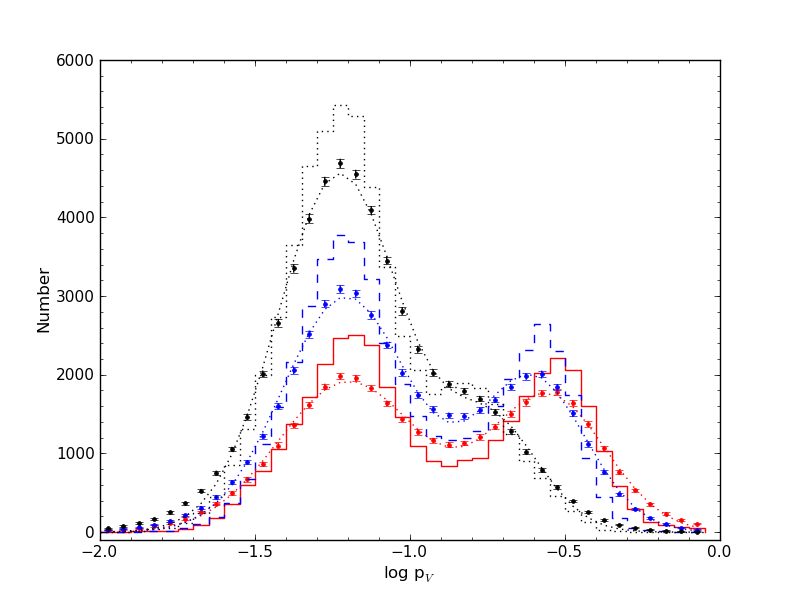}
\protect\caption{The same as Fig~\ref{fig.albAll}, but now including a
  simple model for the objects discovered by NEOWISE that have
  no optical followup photometry.}
\label{fig.albwithlost}
\end{center}
\end{figure}

We show in Fig~\ref{fig.discalb} the albedo distribution of the
NEOWISE-discovered MBAs with optical photometry for the inner-,
middle-, and outer-belt subpopulations.  Even though the optical
followup will be biased toward favoring higher albedo objects, the
NEOWISE discoveries are dominated by low albedos, as these are the
objects that were initially missed by the ground-based optical
surveys.  All three distributions can be described by a single
Gaussian function with mean albedo between $0.05<\mu<0.06$ and
$0.02<\sigma<0.03$, consistent with the values found for the dark
complex in the whole population above.  We can use the albedos of the
discovered objects and the roughly equal number of objects in each
subgroup as an initial attempt to constrain the effect of these lost
objects on the greater albedo distribution.  We show in
Fig~\ref{fig.albwithlost} a revised albedo distribution including this
toy model for the albedos of the $\sim15,000$ Main Belt asteroids
without followup photometry, based on the albedo distribution of the
NEOWISE-discovered objects.  As expected the primary effect is to
increase the relative abundance of dark objects in each region of the
Belt.

We note that the more detailed way to properly account for the
uncertainties introduced by objects without follow-up and objects for
which reliable orbits could not be determined is through careful
modeling of survey biases.  A debiased study of the Main Belt
asteroids will be subject of future work, and will allow us to
determine the true size and albedo distributions of these objects.

\section{Dynamically grouped albedos in the Main Belt}

\begin{figure}[ht]
\begin{center}
\includegraphics[scale=0.8]{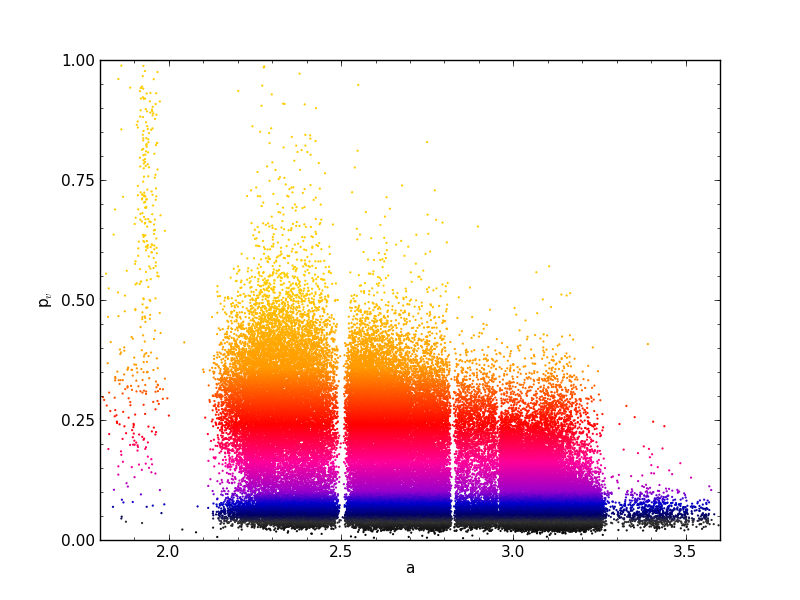} 
\protect\caption{Asteroid albedo vs. semimajor axis.  Colors denoted here are split evenly in $\log(p_V)$ space and are used to denote albedo in subsequent plots.}
\label{fig.apColor}
\end{center}
\end{figure}

We also investigate the distribution of albedo as a function of
orbital parameters, in particular semimajor axis ($a$), eccentricity
($e$) and inclination ($i$).  Figure~\ref{fig.apColor} shows the
distribution of albedos as a function of semimajor axis.  We have
color-coded the points by albedo using a ``weather-map'' palette
divided evenly in $\log(p_V)$ space, and use this same color code for
all subsequent plots.  Dark colors (black, grey, dark blue, blue)
indicate objects in the low albedo complex, while brighter colors
(magenta, red, orange, yellow) indicate members of the high albedo
complex.  Objects colored yellow have very high albedos and are
concentrated in the Hungaria region and inner-Main Belt near the orbit
of (4) Vesta.  These largest albedos found are likely artifacts of
using $G=0.15$ to calculate the $H$ value: while the thermal models
for the diameters of these objects show no errors, using the
literature $H$ value forces anomalously high albedos.  We investigate
the use of different $H$ and $G$ values for these highest albedo
objects in a future publication.

\begin{figure}[ht]
\begin{center}
\includegraphics[scale=0.8]{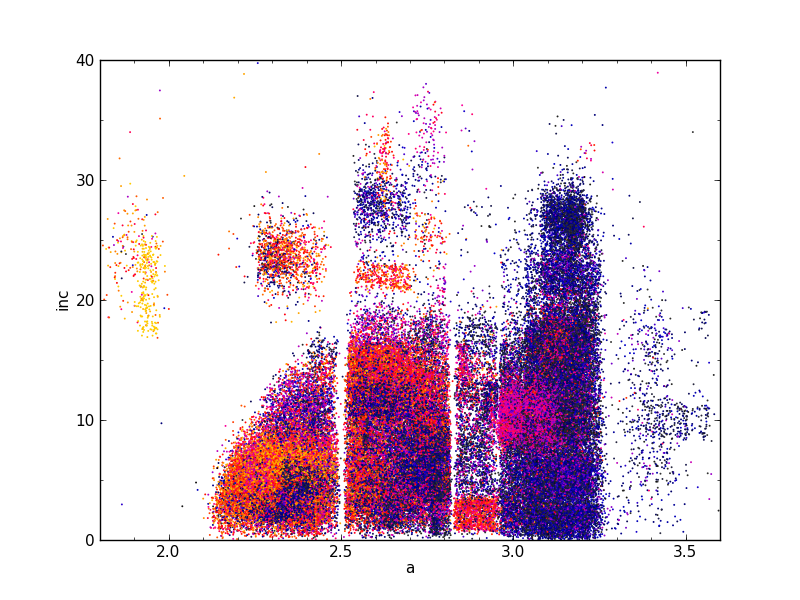} 
\protect\caption{Orbital inclination vs. semimajor axis.  Colors are the same as Fig~\ref{fig.apColor}.}
\label{fig.aip}
\end{center}
\end{figure}

\begin{figure}[ht]
\begin{center}
\includegraphics[scale=0.8]{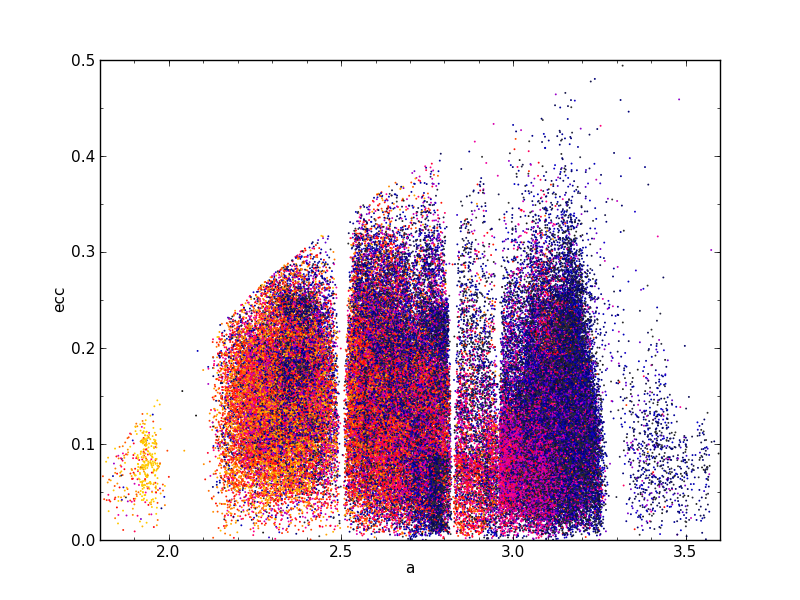} 
\protect\caption{Orbital eccentricity vs. semimajor axis.  Colors are the same as Fig~\ref{fig.apColor}.}
\label{fig.aep}
\end{center}
\end{figure}

We show in Figures~\ref{fig.aip} and \ref{fig.aep} the plot of semimajor
axis vs. inclination and eccentricity, respectively, using the same
colors denoted above.  Asteroid albedos are not homogeneously
distributed throughout the Main Belt, but rather are clumped in a-e-i
space, correlating with the positions of known asteroid families
\citep{nesvorny06}.  MBAs have been previously shown to cluster in
color-space \citep{ivezic02,parker08} and show similar reflectance
spectra \citep[e.g.][etc.]{binzel93,cellino01} indicative of the
common origin of members of families, which are the result of a
catastrophic breakup of a single parent body \citep{hirayama1918}.
The clustering of albedos is a further confirmation of this origin.  A
future paper in this series will investigate the use of albedo as an
added criterion to the orbital parameters typically used to determine
family membership.

\clearpage

\section{IR reflectance}
\label{sec.ratio}

The reflectance spectrum for most asteroids (but not all,
e.g. B-types) show an increasing value as the wavelength moves from
the visible to near-infrared (NIR) region of the spectrum
\citep[e.g.][etc.]{demeo09}.  If these trends continue into the W1 and
W2 bandpasses we would expect the reflectance observed there to be
higher than observed in the optical.  We assume in our thermal model
that the reflectance in W1 and W2 is the same; depending on the
location and depth of absorption bands this may not be universally
true, but this assumption provides a generic constraint from which we
can identify interesting objects that do not follow this assumption.
Reflectance depends on both the albedo and the $G$ slope parameter,
both of which cannot be assumed to be wavelength independent.  We do
not have sufficient phase coverage to fit $G_{NIR}$ and thus
disentangle its effect from that of $p_{NIR}$, so we present only the
NIR reflectance ratio for objects in the Main Belt with sufficient
signal in W1 and/or W2 to be able to fit this value, a total of $4194$
objects.

\begin{figure}[ht]
\begin{center}
\includegraphics[scale=0.75]{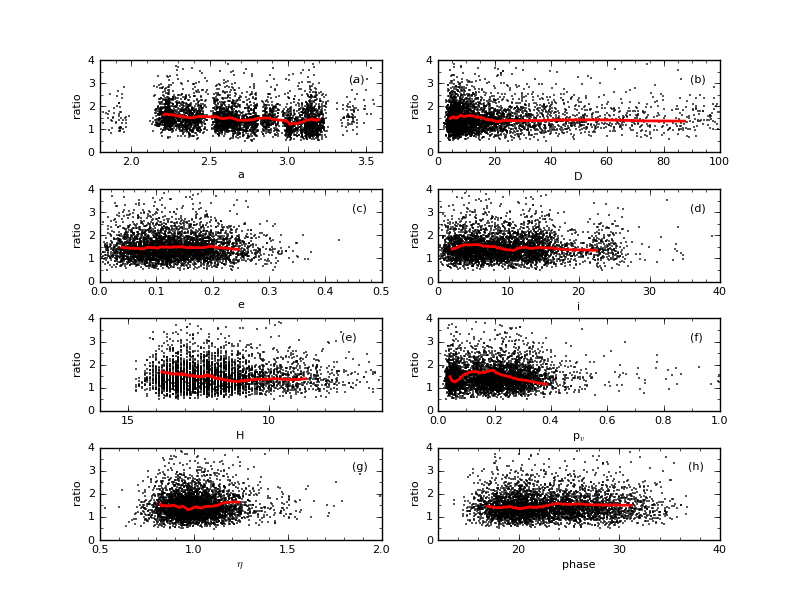} \protect\caption{NIR
  reflectance ratio for all objects with fitted values, compared to
  (a) semi-major axis, (b) diameter, (c) eccentricity, (d)
  inclination, (e) absolute magnitude, (f) albedo, (g) beaming
  parameter and (h) phase.  The thick red line shows the running
  average for $400$ object-wide bins stepped by $40$ objects.  The
  picket-fence effect in the absolute magnitude is an artifact of the
  $0.1~$mag accuracy of $H$ for most objects.}
\label{fig.ratioscat}
\end{center}
\end{figure}

We show in Fig~\ref{fig.ratioscat} plots of the NIR reflectance ratio
compared to a range of physical and orbital parameters, as well as
running averages for those distributions.  All objects were detected
in thermal emission, however only objects with sufficient reflected
light were able to provide fitted NIR reflectance ratios.  As such
objects with higher IR albedos will be more likely to have a fitted
reflectance ratio (this is comparable to the biases inherent in
optical surveys).  Evidence of this is apparent in
Fig~\ref{fig.ratioscat}e, where the running average of the ratio
climbs for objects with fainter $H$ values.  We note that as the
visible albedo $p_V$ is intricately linked with the NIR/Vis ratio it
cannot be considered an independent variable.  The structure seen in
the running average in Fig~\ref{fig.ratioscat}f is expected to be
heavily influenced by the debiasing of the Main Belt population
currently being undertaken.  Future work \citep{mainzer11tax} will
explore the connection between taxonomic types derived from
spectroscopy and photometry and the IR reflectance ratio found in the
WISE data.

\begin{figure}[ht]
\begin{center}
\includegraphics[scale=0.75]{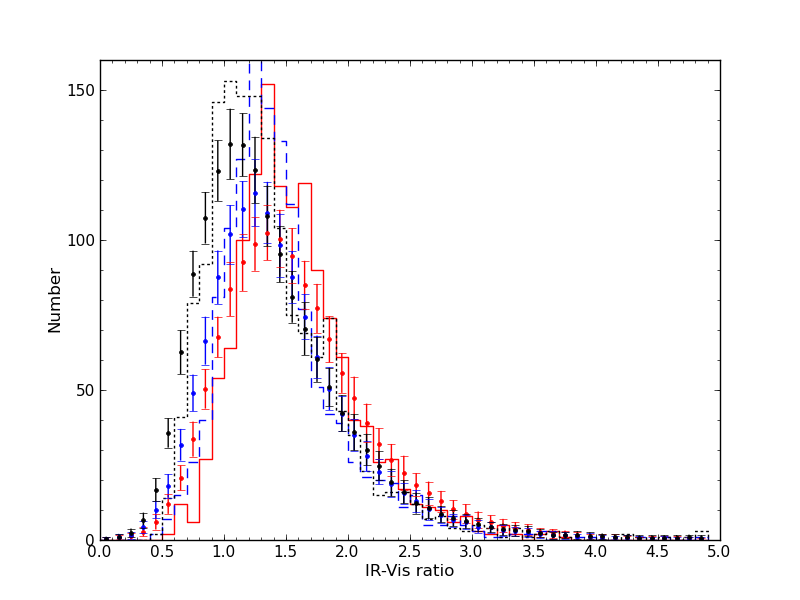}
\protect\caption{Distribution of NIR reflectance ratio for the IMB
  (red solid), MMB (blue dashed) and OMB (black dotted) populations.
  Shown as points are the mean distribution and associated errors from
  Monte Carlo simulations of each ratio.}
\label{fig.ratiodist}
\end{center}
\end{figure}

In Fig~\ref{fig.ratiodist} we show the raw differential
distribution of NIR reflectance ratios for the IMB, MMB, and OMB
asteroids with fitted ratios.  Also shown are the mean distribution
and associated errors derived from a 100-trial Monte Carlo simulation
of all measured reflectance ratios.  All three populations show a peak
between $1.2<$ratio$<1.4$ while the running average vs. semi-major
axis varies across the Main Belt from $1.3<$ratio$<1.6$.  This is
consistent with the values found for the NEOs by \citet{mainzer11neo}.
For objects without fitted NIR reflectance ratios, we use the mean of
all fitted values of $1.5$ and an error bar based on the associated
standard deviation of $\sigma_{ratio}=0.5$ for modeling purposes.

\section{Asteroid families}

Asteroid families were first identified as groups of objects that
clustered tightly in orbital element-space by \citet{hirayama1918}
nearly a century ago.  Subsequent work has confirmed that families
originate from the catastrophic breakup of a single parent asteroid
after an impact \citep[see][for a recent review of the current state
  of the field]{cellino09}.  This single mineralogical origin causes
families to cluster tightly not only when comparing orbital elements
but also when investigating colors \citep{ivezic02,parker08} and
reflectance spectra \citep[e.g.][etc.]{binzel93,cellino01}.  The SFD
of asteroid family members can also act as a tracer of the physical
properties of the original parent body and can even be used to
constrain the impact velocity and angle \citep{durda07}.  However, a
major deficiency in models to date has been the lack of measured
diameters for the family members, forcing these values to be assumed
based on the apparent magnitude of the object.  Albedo measurements of
the largest bodies in a family are often available from the IRAS
data set \citep{tedesco02} and can be used to assume an albedo for all
family members, but this can add a significant and systematic error to
the diameters used, especially in the cases where families may be
mixed or where the largest body in a family may not be associated with
the other members \citep[e.g.][]{cellino01}.

There are a number of methods that can be used to determine which
asteroids are members of a given family.  The Hierarchical Clustering
Method \citep[HCM,][]{zappala90}, a commonly used technique, takes the
differences in velocities between the proper orbital elements of
objects to reveal dynamical associations.  \citet{nesvornyPDS} used
this method to identify $55$ families out of $293,368$ MBAs with low
inclinations.  We use these $55$ families as the baseline for our
analysis, selecting those objects that appear both in that list and in
the WISE observations.  As all these objects were discovered by
optical surveys, there will be an inherent bias in the albedos
favoring brighter family members.  Future work will address this bias
and explore the use of albedo in conjunction with dynamical orbital
properties to identify members of asteroid families and to reject
interloper objects.

Of the $55$ families identified by \citet{nesvornyPDS} we find that
$46$ have more than $20$ members observed during the cryogenic WISE
mission.  Due to the limitations of proper orbital element
calculations, high inclination objects are not included in the AstDys
list\footnote{{\it http://hamilton.dm.unipi.it/astdys/index.php}} of proper
orbital elements \citep{milani94}.  As such, high inclination families
are likewise not represented in the list of family members.  We have,
however, included in our analysis the Pallas family, identified by
Nesvorny (private communication) through the same methods as the $55$
published families.  We also include the asteroids located in the
Hungaria region; while not canonically included in the list of
dynamical families, recent work by \citet{warner09b} and
\citet{milani10} support the classification of objects near Hungaria
in orbital space as a true dynamical family.  These two added groups
bring our total considered population up to $48$ families.

\begin{figure}[ht]
\begin{center}
\includegraphics[scale=0.8]{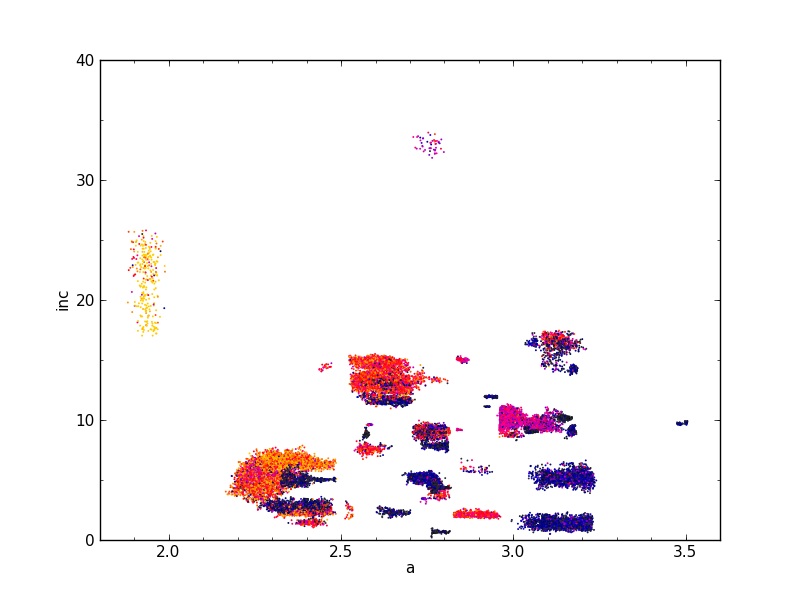} 
\protect\caption{Proper orbital inclination vs. proper semimajor axis for asteroid families.  Colors are the same as Fig~\ref{fig.apColor}.}
\label{fig.aipFam}
\end{center}
\end{figure}

\begin{figure}[ht]
\begin{center}
\includegraphics[scale=0.8]{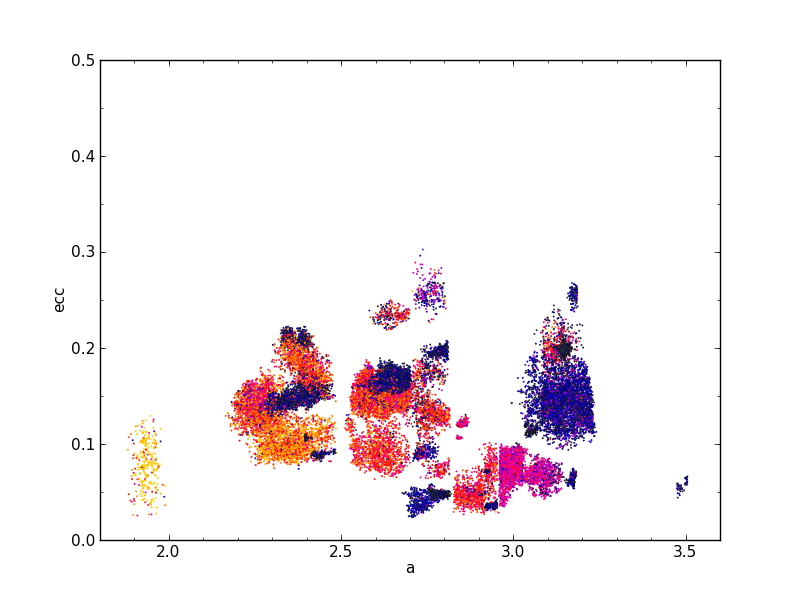} 
\protect\caption{Proper orbital eccentricity vs. proper semimajor axis for asteroid families.  Colors are the same as Fig~\ref{fig.apColor}.}
\label{fig.aepFam}
\end{center}
\end{figure}

We show in Fig~\ref{fig.aipFam} the proper inclination against the
proper semimajor axis for all objects identified as a member of one of
the families considered here.  Note that for the Hungaria and Pallas
families proper orbital elements were not available and so osculating
elements were used.  Similarly, we show in Fig~\ref{fig.aepFam} the
proper eccentricity against the proper semimajor axis.  In both plots,
we use the same color scheme as shown in Fig~\ref{fig.apColor}.  It is
quite apparent in these plots that families characteristically have
uniform albedos, however there are notable exceptions.

\clearpage

\begin{figure}[ht]
\begin{center}
\includegraphics[scale=0.78]{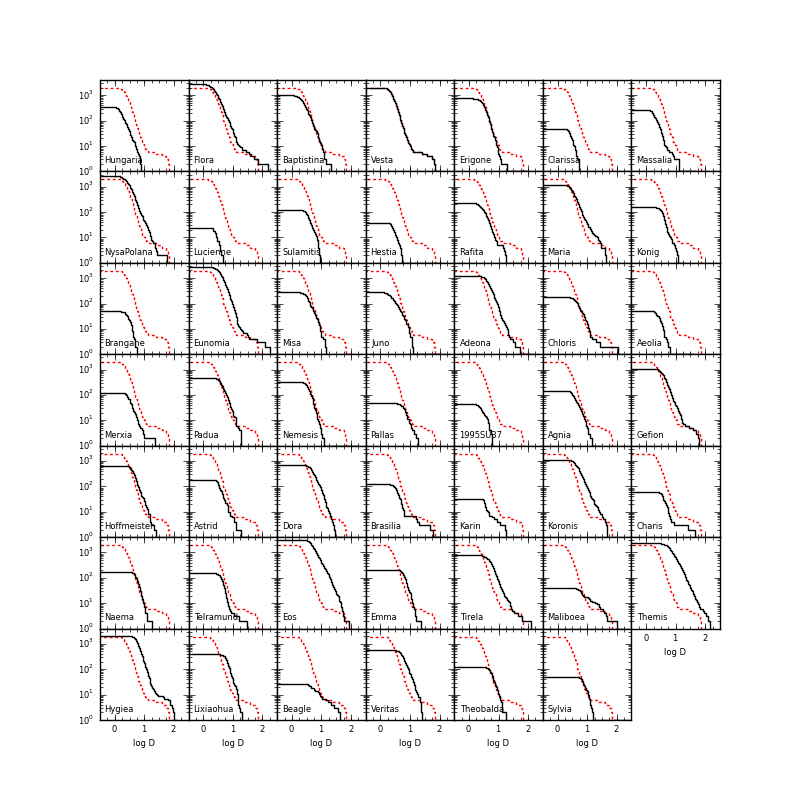}
\protect\caption{Cumulative raw size-frequency distribution for
  each asteroid family considered here.  The solid black line shows
  the family indicated by the name in each plot.  The dotted red line
  shows the PRSFD for the Vesta family is all plots, for ease of
  comparison.}
\label{fig.dCumFam}
\end{center}
\end{figure}

We show in Fig~\ref{fig.dCumFam} the cumulative PRSFD for each of the
families observed during the WISE survey, as well as Monte Carlo
simulations of all distributions with appropriate error bars.  Without
debiasing, the PRSFD cannot be assumed to represent the true size
distribution of the entire family population, as neither the
NEOWISE-inherent biases nor the biases in family selection have been
accounted for.  Approximately $25\%$ of these families show kinks at
the large end of the distribution inconsistent with a simple
power-law.  As small number statistics dominate families especially at
the largest sizes, and because WISE did not survey the entire Main
Belt before the exhaustion of cryogen, precise debiasing is required
to confidently measure the shape of the true SFD, especially at the
largest sizes for each family.

\begin{figure}[ht]
\begin{center}
\includegraphics[scale=0.55]{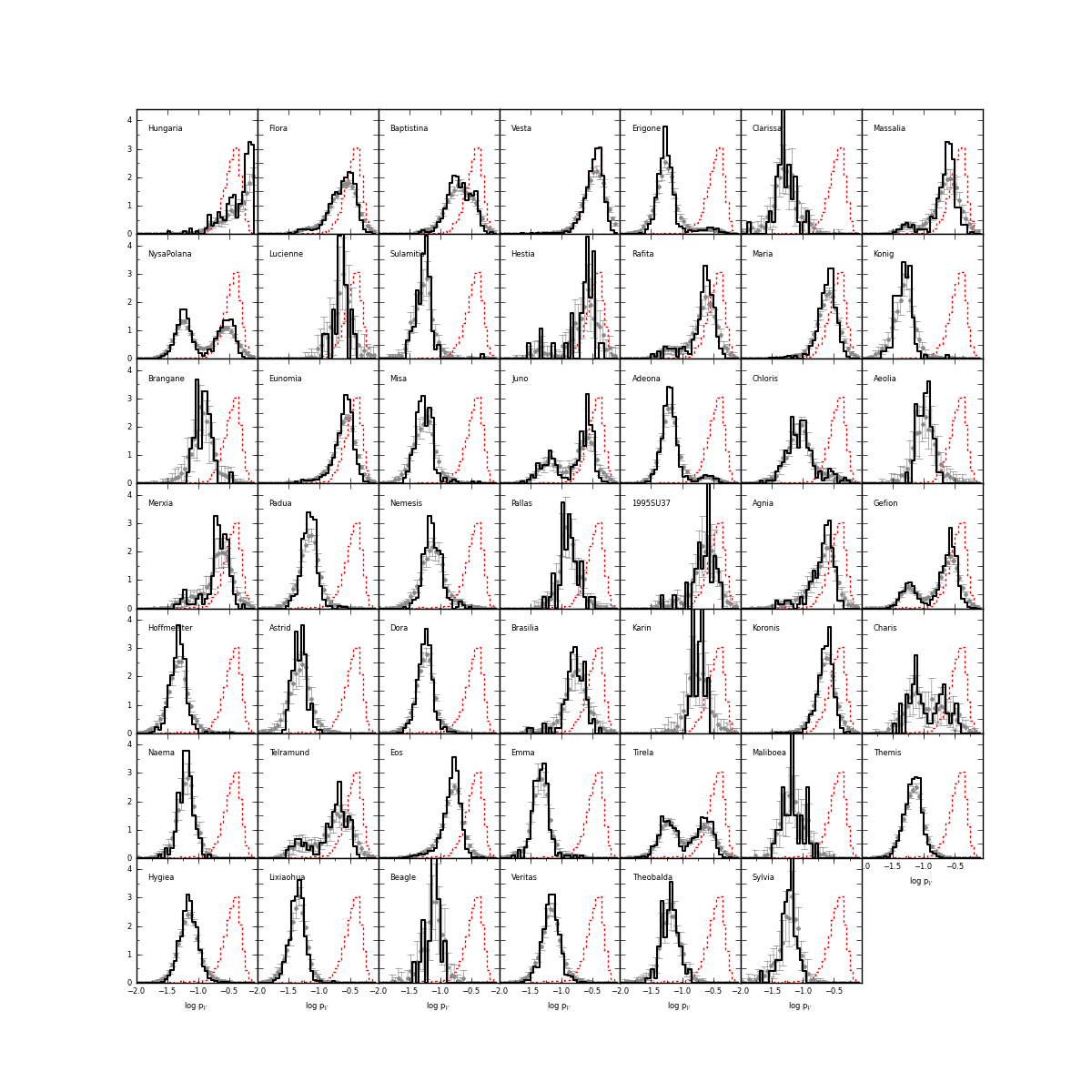}
\protect\caption{The same as Fig~\ref{fig.dCumFam} but for the PRAD.
  Monte Carlo simulations of the distributions are plotted as grey
  points with error bars.  The distributions have been normalized to
  unit area for easier comparison.}
\label{fig.albFam}
\end{center}
\end{figure}

In Fig~\ref{fig.albFam} we show the normalized PRAD for each family,
along with the Monte Carlo results for each distribution.  As was
evident in Fig~\ref{fig.aipFam} and Fig~\ref{fig.aepFam} most families
show a single-peaked albedo distribution, however as mentioned above
debiasing will be critical to proper interpretation of these
distributions.  About $15\%$ of families include a small population of
objects with non-characteristic albedos; these may be the result of an
improper association of background objects into the family.  However,
another $10\%$ of families show significant mixing between two albedo
types that cannot be solely due to the intrusion of a few background
objects.  In particular the Nysa-Polana and Tirela families show
nearly parity between the low and high albedo objects in the
preliminary raw distributions.  

Among the families with sufficient data for study were the Karin and
Koronis families.  The Karin family is believed to be a very young
family, with an age of $\sim5.8~$Myr, that formed from the breakup of
a member of the much older ($2-3~$Gyr) Koronis family
\citep{nesvorny06karin}.  We fit a Gaussian to the observed family
albedo distributions.  We find that the Karin family has a lower mean
albedo from this fit ($p_{V-Karin}=0.18\pm0.05$) than the Koronis
family ($p_{V-Koronis}=0.24\pm0.05$) where the error bars represent
the width of the best-fit Gaussian.  \citet{chapman04} gives an
overview of space weathering effects, a process that is generally
thought to darken and redden surfaces of atmosphereless bodies in the
Solar system.  Our result is in apparent contradiction with this
hypothesis under the assumption that the compositions of both families
are identical.  It is possible that variations in the composition, the
presence of interlopers, or differentiation of the proto-Koronis
parent body could result in this observation.  This analysis will be
strengthened by the identification of more NEOWISE-observed family
members (there were $31$ identified members of the Karin family and
$1079$ members of the Koronis family in this data set) and the
acquisition of additional compositional information.

We also observed $984$ asteroids that were identified as members of
the Baptistina family.  \citet{bottke07} postulate that a fragment
from the breakup of the Baptistina family was the impactor responsible
for the K/T mass extinction event.  However, these authors assumed an
albedo for the family members of $0.05$.  We find that the best
fitting Gaussian to the Baptistina family member albedos has a mean of
$p_{V-Baptistina}=0.21^{+0.13}_{-0.08}$ where the error bar indicates
the width of the Gaussian distribution.  The method of age
determination used by \citet{bottke07} depends on the albedo assumed,
and the calculated age $T$ is proportional to albedo following $T
\propto p_V^{-0.5}$.  Thus our measured albedo results in an age for
the breakup approximately $\sim2$ times younger than that found with
the lower assumed albedo (from $160~$Myr to $\sim80~$Myr), reducing
the likelihood that the Baptistina breakup generated the K/T impactor
\citep[c.f.][]{reddy09}.

Also of interest are the families that show characteristic albedos
distinct from the distribution observed in the Main Belt in
Fig~\ref{fig.albAll}.  For example, the Eos and Aeolia families have
characteristic albedos that fall in between the two peaks of the MBA
albedos.  Standing out as an outlier is the Hungaria family: while
showing a fairly strong coherence in albedo within the family, that
characteristic albedo is incredibly high (see below).  Future work
will investigate whether variations in the $G$ slope parameter used to
compute the $H$ absolute magnitude (and thus the reflected-light
albedo) from the assumed value of $G=0.15$ typically used could
account for the very large albedos calculated in these preliminary
results.

\clearpage

\section{Unusual objects}
\label{sec.unusual}
A small fraction of our objects had fits that did not conform to the
general trends seen for the population as a whole.  These highly
unusual fits may be caused by strange physical parameters, incorrect
associated properties (e.g. $H$ mag, orbital elements, etc), or a
break-down of the NEATM model.  In any case, these objects warrant
further inspection.

The most obvious candidates for this category of unusual fits are the
asteroids with very high visible geometric albedos ($p_V\ge0.70$).  We
find $193$ objects in our survey have computed albedos that fall into
this range, mostly contained within the region of orbital element
space occupied by the Hungaria and Vesta families (out of $343$
objects found in the Hungaria region and $1938$ in the Vesta family),
implying a possible mineralogical origin.

The asteroid (434) Hungaria, the lowest numbered member of its
namesake region, has a relatively large albedo of $p_V=0.46$ which can
be explained by a composition dominated by the iron-poor mineral
enstatite however some of the observed spectral features may require
contamination from a darker, external source \citep{kelley02}.  This
may indicate that albedos can range larger than $p_V\sim0.5$, but
albedos significantly higher than this are likely suspect.

Visual inspection of the thermal model fits of these objects shows
that these high albedos are not due to a failure of the thermal model.
For those objects where IR reflectance factor could also be fitted, we
find values typically with reflectance ratio$\le1$, while the
distribution of beaming parameters for these objects is similar to
that of the general Main Belt population.  If the $H$ or $G$ values
for these objects were very far from the true values this could result
in the unusual calculated albedos.

By increasing the error bar we assume for the $H$ value and setting
the IR reflectance ratio to a constant value of ratio$=1.5$ (the
average for MBAs) we are able to find good fits of objects with
reflected light in W1 and/or W2.  Under this assumption, these objects
all return fits with albedos comparable to that of (434) Hungaria.
However all these cases require the $H$ magnitudes to be $0.4-1.0~$mag
fainter than the values given by the Minor Planet Center.

A mis-identification of $H$ could be a symptom of an improper
assumption for the $G$ slope parameter.  In particular if $G$ should
be much larger (e.g. equal to or greater than the value for (44) Nysa
of $G=0.46$\footnote{as given by the Small Body Database: {\it
    http://ssd.jpl.nasa.gov/sbdb.cgi}} this would result in an $H$
value that was too bright by $\sim0.4~$mag or more, which would thus
give an albedo that was a factor of $\sim1.4$ too large.  This can
account for some, but not all, of our improbably high albedos.  Large
light curve amplitudes may also contribute to an absolute magnitude
that is too bright.  Future work will address these objects in detail.

In addition to the high albedo objects, we find a very small number of
objects with beaming parameters at or close to the theoretical limit
of $\eta=\pi$.  While some of these fits can be rejected upon visual
inspection e.g. because of a single bad point in W4 dragging the fit
to higher beaming parameter, at least $6$ objects appear to be
legitimate fits with beaming parameters indicative of
latitudinally-isothermal surfaces, though none of the fits show any
significant lightcurve variations.  Further investigation will be
critical to determine if they show rapid rotation or very high thermal
inertia needed to explain this beaming parameter.

Approximately $10,000$ MBAs were observed by WISE at two different
epochs.  We initially treated each epoch separately for fitting
purposes, and in the majority of cases the fits were within the
expected error of each other.  For those that weren't, we recomputed
the best fitting model using both epochs together and forcing the
physical parameters to be identical.  We find $36$ objects for which
the two-epoch fits could not produce a good fit at one or both epochs.
While some of these objects may have very long rotation
periods($P>10~$days) and thus different projected diameters between
epochs, others do not appear to show any light curve variation during
our observations and may be cases of objects showing a significant
difference between the temperatures of the morning and afternoon
hemispheres of the body.

\begin{figure}[ht]
\begin{center}
\includegraphics[scale=0.6]{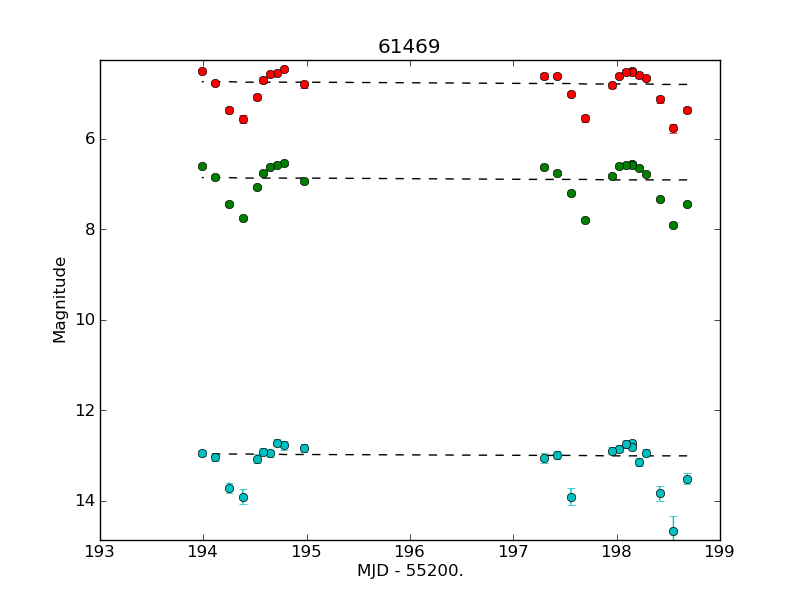} \protect\caption{Light
  curve for asteroid (61469), with magnitudes for $W4$, $W3$ and $W2$
  shown in red, green and cyan (top to bottom) respectively.  The
  dashed line shows the modeled magnitudes for the best-fitting
  sphere.  The period of this asteroid is approximately $\sim40~$hr
  assuming a double-peaked lightcurve.}
\label{fig.LC}
\end{center}
\end{figure}

We also observe $151$ objects with peak-to-trough light curve
variations larger than $1.5~$mag in W3 and average magnitude
measurement error smaller than $\sigma_{W3}<0.2$ after removing
spurious measurements.  As these are not fitted amplitudes they
represent a minimum for the light curve amplitude for the body
observed.  An example of one of these objects, (61469), is shown in
Fig~\ref{fig.LC}.  There are also many objects with amplitudes smaller
than this cutoff with readily apparent rotational effects, too.  A
future work will investigate specific light curves to determine the
period and amplitude of the objects.

\section{Conclusions}
We have presented an initial analysis of Main Belt asteroids detected
by NEOWISE during the cryogenic portion of the WISE mission.  With
infrared fluxes of sufficient quality to determine diameters and
albedos for $129,750$ MBAs, we show the power and great potential
contained in this dataset.  These data allow us to probe the
composition, structure, and history of the Main Belt in ways that were
previously impossible.

For objects with thermal emission detected in two or more bands we
allowed the beaming parameter to vary.  We find a mean beaming
parameter of $\eta\sim1.0$ however we do see evidence of a phase
dependence for the beaming parameter, ranging from $\eta\sim0.94$ for
low phases to $\eta\sim1.14$ for higher phases (within the Main Belt).
The best fit linear relation between beaming and phase is $\eta=0.79 +
0.011 \alpha$ which is a much shallower relation than seen previously
in \citet{wolters09}, but consistent with \citet{mainzer11neo} which
included the near Earth objects with fitted beaming parameters as well.

As was observed in the IRAS data
\citep[e.g.][]{tedesco89,tedesco02,tedesco05} the albedos of Main Belt
asteroids are strongly bimodal: a bright complex ($p_V\sim0.25$) and
the dark complex ($p_V\sim0.06$).  We find both peaks to be well
described by Gaussian distributions in log-albedo space.

We find that the reflectance of asteroids in the W1 and W2 bandpasses
is typically larger than the albedo found in visible light.  The best
fit ratio of reflectance ranges from $\sim1.6$ in the IMB to $\sim1.3$
in the OMB however the spread of values is large, and the final
distribution will depend strongly on the debiasing.

We identify albedo clusters in $a-e-i$ space corresponding to the
locations of asteroid families.  Albedo is another coherent property
of dynamical families in addition to orbit \citep{hirayama1918}, color
\citep{ivezic02} and reflectance spectrum
\citep[e.g.][etc.]{binzel93,cellino01}.  Albedo can also be used to
trace the halos of similar objects that surround some families
\citep[e.g. Vesta, Eos, etc.,][]{parker08} that may be evidence of a
collisional breakup very early in the age of the Solar System.  Using
asteroids previously identified though HCM techniques to be members of
collisional families we show that most, but not all, families have a
characteristic albedo.

Critical to any interpretation of the observations presented here is a
careful accounting of the biases in both the WISE survey data as well
as the optical data used to derive albedos.  We are currently
undertaking an extensive debiasing campaign with the goal of producing
unbiased size and albedo distributions for the Main Belt.  This will
be the subject of the next paper in this series.

\section*{Acknowledgments}

J.R.M. was supported by an appointment to the NASA Postdoctoral Program
at JPL, administered by Oak Ridge Associated Universities through a
contract with NASA.  J.R.M. thanks M. Delb{\'o} and M. Mueller for
providing access to their thermal modeling code which was helpful in
early test cases.  This publication makes use of data products from
the Wide-field Infrared Survey Explorer, which is a joint project of
the University of California, Los Angeles, and the Jet Propulsion
Laboratory/California Institute of Technology, funded by the National
Aeronautics and Space Administration.  This publication also makes use
of data products from NEOWISE, which is a project of the Jet
Propulsion Laboratory/California Institute of Technology, funded by
the Planetary Science Division of the National Aeronautics and Space
Administration.  This research has made use of the NASA/IPAC Infrared
Science Archive, which is operated by the Jet Propulsion Laboratory,
California Institute of Technology, under contract with the National
Aeronautics and Space Administration.  We thank the worldwide
community of dedicated amateur and professional astronomers devoted to
minor planet follow-up observations.  We are deeply grateful for the
outstanding contributions of all members of the WISE and NEOWISE
teams.

\end{document}